\def\n{\noindent}

\def\be{\begin{equation}}
\def\ee{\end{equation}}

\def\eq{\enskip =\enskip}

  \def\ket{\vert \vert  \{ \emptyset \} \rangle}
  \def\ket2{\vert \vert \otimes \{ R \} \rangle}

\def\.#1{\mathaccent 95#1}
\def\^#1{\mathaccent 94 #1}
\def\~#1{\mathaccent "7E #1}

\def\eq{\enskip =\enskip}

  \def\ket{\vert \vert  \{ \emptyset \} \rangle}
  \def\ket2{\vert \vert \otimes \{ R \} \rangle}

\def\PR{{\it Phys. Rev.}}
\def\JPC{{\it J. Phys. C:Solid State}}
\def\JPCM{{\it J. Phys. Condens Matter}}
\def\q{{\mathbf q}}

\def\g{{\mathbf g}}
\def\D{{\mathbf D}}
\def\mbf{\mathbf }
\def\A{{\mathbf A}}
\def\F{{\mathbf F}}
\def\B{{\mathbf B}}
\def\W{{\mathbf W}}
\def\wt{\widetilde }

\documentclass[twocolumn,showpacs,preprintnumbers,amsmath,amssymb]{revtex4}
\usepackage{graphicx}% Include figure files
\usepackage{dcolumn}% Align table columns on decimal point
\usepackage{bm}% bold math

\begin{document}
\date{\today}
%\preprint{SNB/CMP/04-8}

\title[Inelastic neutron scattering in disordered systems]{Inelastic neutron scattering in random binary alloys~: an augmented space approach}

\author{Aftab Alam \footnote{corresponding author}\footnote{email : alam@bose.res.in} and Abhijit Mookerjee }

\affiliation{ S.N. Bose National Centre for Basic Sciences,
 JD Block, Sector III, Salt Lake City, Kolkata 700098,
India}

\begin{abstract}
Combining the augmented space representation for phonons with a generalized version of Yonezawa-Matsubara diagrammatic technique, we have set up a formalism to seperate the coherent and incoherent part of the total intensity of thermal neutron scattering from disordered alloys. This is done {\sl exacly} without taking any recourse to mean-field like approximation (as done previously). The formalism includes disorder in masses, force constants and scattering lengths. Implementation of the formalism to realistic situations is performed by an augmented space 
{\sl Block recursion} which calculates entire Green matrix and self energy matrix which in turn is needed to evaluate the coherent and incoherent intensities. we apply the formalism to $Ni_{55}Pd_{45}$ and $Ni_{50}Pt_{50}$ alloys. Numerical results on coherent and incoherent scattering cross sections are presented along the highest symmetry directions. Finally the incoherent intensities are compared with the CPA and also with experiments.

\end{abstract}

\maketitle

\section{Introduction}

Phonon excitations in alloys provide a particularly severe testing arena for any theory of elementary excitations in disordered systems. The dispersion relations, line widths etc. for phonons can be measured directly through {\it coherent inelastic neutron scattering}. Information about the phonon density of states can be extracted from the {\it incoherent inelastic neutron scattering cross sections}. Over the past few years, numerous experimental studies \cite{Tsunoda,Svensson,Smith,kambrock,Nicklow} of the lattice dynamics of disordered systems have been carried out providing deep insight into the nature of their elementary excitations. However the theoritical counter part is still unsatisfactory. The theory of scattering of thermal neutrons with perfect crystals has been set up on a rigorous basis \cite{vH,sjo,lm}.However, we have to make the same statement for disordered alloys with more care. This is because the theory of neutron scattering in random alloys require two basic inputs : first is the formulation of the problem and the second it's actual numerical implementation in realistic situations. As far as the formulation part is concerned, several authors have attempted the problem with different approaches. However their actual numerical implementation in realistic systems still remains sketchy beyond the simple single site mean field theories. Quite some time ago, Nowak and Dederichs \cite{nd} discussed the  separation of  the coherent and incoherent parts of the total inelastic scattering intensity by using the Yonezawa-Matsubara diagramatic technique within the single site coherent potential approximation (CPA).According to their approach the incoherent part of the total scattering intensity is the sum of all irreducible diagrams containing only short ranged correlations. The coherent part on the other hand may be expressed as a product of the usual configuration averaged Green function and the square of an effective scattering length, which is itself given by the irreducible diagrams closely related to those for the self energy.It has been known for some time that the single site CPA cannot adequately deal with intrinsic off-diagonal disorder of the force constants in the problem of phonon excitations in random alloys. This was evidenced in the inability of the single site CPA to explain experimental lifetime data on NiPt \cite{Tsunoda}. Nor can it adequately deal with the correlated diagonal and off-diagonal disorder induced by the force constant sum rule. Several successful attempts have been made to go beyond the CPA. These include, among others, approximations based on the augmented space formalism (\cite{Mookerjee}) : the travelling cluster approximation (TCA) \cite{kapmos,kapgray,mills,KLGD}, the Cluster-CPA \cite{mks,msc}, the itinerant cluster approximation (ICPA) \cite{glc} and the augmented space recursion (ASR) \cite{alam}. 

In this  communication we shall tackle a two-fold problem : one of formulation and the other of implementation in real alloy systems. We shall use a scattering technique also based on the augmented space formalism \cite{mook75a,mook75c} to suggest how to separate the coherent and incoherent parts of the total inelastic scattering cross-section for a disordered binary alloy in a way mirroring the ideas of Nowak and Dederichs, but  this will be done without taking any recourse to mean-field like approximations. For implementation in real alloy systems, we shall suggest the ASR for the evaluation of scattering cross sections. But instead of doing an ordinary recursion, we shall perform a {\it Block recursion} in order to calculate the off diagonal entries of the Green matrix, since the expressions for the scattering cross sections in our formalism require the contribution of off-diagonal Green's function. The approximation introduced within this formalism will maintain the essential analytic properties of the Green function, deal with off-diagonal disorder and the sum rule without any further simplifications or assumptions and encompass environmental effects over an extended neighbourhood. This is the major contribution of this work.

The rest of the paper is organised as follows. In Sec. II, we describe the theory, introducing the augmented space representation and it's use in constructing the generalized Yonezawa-Matsubara scattering diagrams for the averaged green function. In Sec. III we derive expressions for important physical quantities such as effective scattering length $W_{eff}({\bf q},w)$, inelastic coherent and incoherent scattering cross sections in terms of the configuration averaged Green matrix $\ll G({\bf q},w) \gg$ and self energy matrix $\Sigma ({\bf q},w)$ of the system. In Sec. IV, we briefly describe the method of {\it Block recursion} for the evaluation of full Green matrix and Self energy matrix. In Sec. V we present our results on $Ni_{55}Pd_{45}$ and $Ni_{50}Pt_{50}$ alloys as test cases and try to compare them with experimental data as far as applicable. Concluding remarks are presented in Sec. VI.

\section{Formalism}

\subsection{The augmented space representation for  phonons}

The augmented space formalism (ASF) has been described in detail in several earlier papers (see \cite{tf}). We shall, for the sake
of completeness,  describe only those features which will be necessary for the implementation of our ideas in this
communication. 
The basic problem in the theory of phonons is to solve a secular equation of the form~: 
 \[ ({\bf M}w^{2} - {\bf D})\ {\bf u}(R,w) = 0 \]  where $u_{\alpha}(R,w)$ is the fourier transform of $u_{\alpha}(R,t)$, the displacement of an atom from its equilibrium position $R$ on the lattice, in the direction ${\alpha} $ at time $t$. {\bf M} is the {\it mass operator}, diagonal in real-space
 and {\bf D} is the {\it dynamical matrix operator} whose tight-binding representations are :
\begin{eqnarray*}
{\bf M} &=& \sum_{R}  m_{R}\ {\delta}_{\alpha \beta} \ P_R\\
{\bf D} &=&  \sum_{R} \Phi_{RR}^{\alpha \beta}\ P_{R} + \sum_{R}\sum_{R^{\prime} \ne R} \Phi_{RR^{\prime}}^{\alpha \beta}\ T_{RR^{\prime}}
\end{eqnarray*}
along with the {\it sum rule}:
\begin{equation}  
\Phi_{RR}^{\alpha \beta} = -\sum_{R^{\prime}\ne R}\Phi_{RR^{\prime}}^{\alpha \beta}
\end{equation}

\noindent Here $P_R$ is the projection operator\ $\vert R\rangle\langle R\vert$\ \ and $T_{RR'}$ is the transfer operator\ $\vert R\rangle\langle R'\vert$\ \ in the Hilbert space ${\cal H}$ spanned by the tight-binding basis $\{\vert R\rangle\}$.
 $R,R^{\prime}$ specify the lattice sites and $\alpha $,$ \beta $ the Cartesian directions. $m_{R}$ is the mass of an atom occupying the position $R$ and $\Phi_{RR^{\prime}}^{\alpha \beta}$ is the force constant tensor. 

\noindent We shall be interested in calculating the displacement-displacement Green matrix ${\mbf G}
(R,R',w^2)$ :

\[ {\mbf G}(R,R',w^2) = \langle R | \left({\bf{M}}w^{2}-{\bf{D}}\right)^{-1} | R' \rangle \]

Let us now consider a binary alloy $ A_{x}B_{y} $ consisting of two kinds of atoms A and B of masses
 $m_A$ and  $m_B$ randomly occupying each lattice sites. We wish to calculate the configuration-averaged
 Green matrix $\ll {\bf G}({R,R'},w^2)\gg$. We shall use the augmented space formalism to do so and indicate the main operational results here.  For further details we  refer the reader to the above monograph \cite{tf}.
The first operation is to represent the random parts of the secular equation in terms of a random
set of local variables $\{ n_R\}$ which are 1 if the site $R$ is occupied by an  A  atom and 0
if it is occupied by B. The probability densities of these variables may be written as :
 
\begin{eqnarray}
 Pr(n_R)& = & x\ \delta (n_{R}-1)\ +\ y\ \delta(n_{R})\nonumber\\
   & =  & (-1/{\pi})\ \Im m\langle {\uparrow}_{R} |\ (n_{R}I-{\it {N_{R}}})^{-1}\ | {\uparrow}_{R}\rangle \label{prob}
\end{eqnarray}

\noindent where $x$ and $y$ are the concentrations of the constituents A and B with $x+y=1$. $N_{R}$ is an operator defined on the configuration-space $\phi_{R}$ of the variable $n_{R}$. This is of rank $2$ and is spanned by the states $\{|{\uparrow_{R}}\rangle, |{\downarrow_{R}}\rangle \}$.  $|{\uparrow_{R}}\rangle$ is a state which indicate that there is an atom A at the site R, while $|{\downarrow_{R}}\rangle$ is for the atom B.

\[ N_{R} = xp_{R}^{\uparrow} + yp_{R}^{\downarrow} + \sqrt{xy}\ {\cal T}^{\uparrow\downarrow}_R \quad\mbox{;}\quad{\cal T}^{\uparrow\downarrow}_R\ =\ {\tau}^{\uparrow \downarrow}_{R} + {\tau}^{\downarrow \uparrow}_{R} \]

\noindent  where $p_R^\uparrow =\vert\uparrow_R\rangle\langle\uparrow_R\vert$ and $p_R^\downarrow=\vert\downarrow_R\rangle\langle\downarrow_R\vert$ are projection operators 
and ${\tau}_R^{\uparrow\downarrow}=\vert\uparrow_R\rangle\langle\downarrow_R\vert$ and
 ${\tau}_R^{\downarrow\uparrow}=\vert\downarrow_R\rangle\langle\uparrow_R\vert$ are transfer operators in the configuration space $\phi_{R}$. 

In terms of random variables $n_{R}$, the mass operator can be written as :
\begin{equation} 
{\bf M}\ =\ \sum_{R}  \left[\rule{0mm}{4mm} m_{B}\ +\ n_{R}\ (\delta m) \right] \delta_{\alpha \beta} \ P_R \quad ;\quad \delta m=m_A-m_B
\end{equation}

\noindent According to the augmented space theorem, in order to obtain the configuration-average we simply replace the random variables $n_R$ by the
corresponding operators $N_R$ associated with its probability density, as in Eqn. (\ref{prob}), and take the matrix element of
the resulting operator between the {\sl reference states}. For a full
mathematical proof the reader is referred to \cite{Mookerjee}.

\[
 n_{R}\longrightarrow N_{R} \ =\  
x\ \tilde{I}\ +\ (y-x)\  p_{R}^{\downarrow} + \sqrt{xy}\ {\cal T}^{\uparrow \downarrow}_R  
\]

\noindent Using the above we get,

%\begin{widetext}
\begin{eqnarray}
\widetilde{\bf M}&=& \A({\mbf m} )\ \widetilde{I}\otimes  I + \B({\mbf m} )\ \sum_{R} p_{R}^\downarrow \otimes P_{R}\ldots \nonumber \\ & &\ldots+ \F({\mathbf m} )\ \sum_{R}\  {\cal T}^{\uparrow\downarrow}_R \otimes P_{R}\nonumber\\ 
\phantom{\widetilde{\bf M}} &=& \ll \widetilde{\mbf M}\gg\  +\  \widetilde{\mbf M}^\prime
\label{mass}
\end{eqnarray}
%\end{widetext}
\noindent where

\[ \left.  \begin{array}{ll}

\A({\mathbf X}) = \ll {\mbf X} \gg\ =(x {\mathbf X}_A+y {\mathbf X}_B) \\
\B({\mathbf X}) = (y-x)\ ({\mathbf X}_A-{\mathbf X}_B) \\
\F({\mathbf X}) = \sqrt{xy} \ ({\mathbf X}_A-{\mathbf X}_B)  \end{array} \right\} \]

\noindent Similarly the random off-diagonal force constants $\Phi_{RR^{\prime}}^{\alpha \beta}$ between the sites $R$ and $R^{\prime}$ can be written as :

\begin{widetext}
\begin{eqnarray}
\Phi_{RR^{\prime}}^{\alpha \beta} &=& \Phi_{AA}^{\alpha \beta} n_{R} n_{R^{\prime}} + \Phi_{BB}^{\alpha \beta} (1-n_{R}) (1-n_{R^{\prime}})\ +\  \Phi_{AB}^{\alpha \beta} \left[\rule{0mm}{4mm}\ n_{R}(1-n_{R^{\prime}}) + n_{R^{\prime}}(1-n_{R})\ \right]\nonumber\\
\phantom{x} \nonumber\\
&=&  \Phi_{BB}^{\alpha \beta}\ +\ \left(\rule{0mm}{4mm}\Phi_{AA}^{\alpha \beta} + \Phi_{BB}^{\alpha \beta} - 2 \Phi_{AB}^{\alpha \beta}\right)\ n_{R} n_{R^{\prime}}\ 
 +\  \left(\rule{0mm}{4mm}\Phi_{AB}^{\alpha \beta} - \Phi_{BB}^{\alpha \beta})\ (n_{R} + n_{R^{\prime}}\right) \nonumber \\
\end{eqnarray}
\end{widetext}

\noindent Let us define the following :

\begin{eqnarray*}
\Phi^{\alpha\beta}_{(1)} &=& x\ \Phi_{AA}^{\alpha \beta} - y\ \Phi_{BB}^{\alpha \beta} + (y-x) \Phi_{AB}^{\alpha\beta}\\
\Phi^{\alpha\beta}_{(2)} &=& \Phi_{AA}^{\alpha \beta} + \Phi_{BB}^{\alpha \beta} - 2 \Phi_{AB}^{\alpha \beta}
\end{eqnarray*}

\noindent In augmented space the off-diagonal force constant matrix becomes an operator :

\begin{widetext}
\begin{eqnarray*}
\widetilde{\mbf D}^{\alpha\beta}_{(off)} &=& \sum_{RR'}\ \left[\rule{0mm}{5mm} \ll \Phi^{\alpha\beta}_{RR'}\gg\ \tilde{I} + 
  \Phi^{\alpha\beta}_{(1)} \left\{ (y-x)\ (p^\downarrow_R
+p^\downarrow_{R'})
\ +\ \sqrt{xy} ({\cal T}^{\uparrow\downarrow}_{R}+{\cal T}^{\uparrow\downarrow}_{R'})\right\}+\ldots\right.\\
& & \left.\ldots + \Phi^{\alpha\beta}_{(2)}\ \left\{ (y-x)^2\ p^\downarrow_R\ p^\downarrow_{R'} + 
\sqrt{xy}(y-x) \left(p^\downarrow_R\ {\cal T}^{\uparrow\downarrow}_{R'} + p^\downarrow_{R'}\ {\cal T}^{\uparrow\downarrow}_{R}
\right) + xy\ {\cal T}^{\uparrow\downarrow}_{R}{\cal T}^{\uparrow\downarrow}_{R'} \right\}\rule{0mm}{5mm} \right]\otimes T_{RR'} \\
\phantom{x}\\
&=& \sum_{RR'} \ll \Phi^{\alpha\beta}_{RR'}\gg I\otimes T_{RR'} +  \sum_{RR'}\ \Psi_{RR'}^{\alpha\beta}\otimes T_{RR'} \\
\end{eqnarray*}
\end{widetext}

\noindent The sum rule gives the diagonal element :
\begin{eqnarray*}
\widetilde{\mbf D}^{\alpha\beta}_{(dia)}& =& -\sum_{R}\left\{\rule{0mm}{1mm}\sum_{R'\ne R} \ll \Phi^{\alpha\beta}_{RR'}\gg \widetilde {I}\right\} \otimes P_R \ldots\\
& &  - \sum_{R}\ \left\{ \sum_{R'\ne R}  \Psi_{RR'}^{\alpha\beta}\right\}\otimes P_R
\end{eqnarray*}

\noindent The total dynamical matrix in the augmented space is :

\begin{eqnarray}
 \widetilde{\mathbf D} &\ =\ & \ll\widetilde{\mbf D}\gg  - \sum_{R}\ \left\{ \sum_{R'\ne R} \Psi_{RR'}^{\alpha\beta}\right\}\otimes P_R \ \ldots\nonumber\\
& & \ldots \ +\ 
 \sum_{RR'}\ \Psi_{RR'}^{\alpha\beta}\otimes T_{RR'} \nonumber\\
 \phantom{\widetilde{\mathbf D}} &\ =\ & \ll\widetilde{\mbf D}\gg\ +\ \widetilde{\mbf D}^\prime  
\label{dm}
\end{eqnarray}

\n The boldface operators are $3\times 3$ matrix representations in the three Cartesian directions.

The augmented space theorem \cite{Mookerjee} now states that the configuration-average of the Green matrix $\ll~{\bf G}({R,R'},w^2)~\gg$ may be written as :
\begin{eqnarray}
\ll {\mbf G}\left({R,R'},w^{2}\right)\gg \phantom{xxxxxxxxxxxxxxxxxx}\nonumber\\
\phantom{xxx}  = \int {\mbf G}\left({R,R'},w^2,\{n_{R}\}\right)\ \prod\ Pr(n_{R})\ dn_{R}\nonumber\\
 \phantom{xxx} =   \langle  \{ \emptyset \}\otimes R|\ \widetilde {\mbf G}(w^{2},\{N_{R}\})| \{\emptyset\}\otimes R'\rangle\phantom{GxxG}\nonumber\\ 
\phantom{x}\nonumber\\
\phantom{xxx}  =  \langle \{ \emptyset \}\otimes R|\ \left(\widetilde{\bf {M}}\  w^{2} -  
 \widetilde{\bf {D}}\right)^{-1}\   |\{\emptyset\} \otimes R'\rangle\phantom{xx}
\label{g0}
\end{eqnarray}
\noindent where $\widetilde{\bf{M}}$ and $\widetilde{\bf {D}}$ are the operators 
which are constructed out of ${\bf{M}}$
 and ${\bf {D}}$ by replacing all the random variables $n_{R}$ (or $n_{R^{\prime}})$ by 
the corresponding operators $N_{R}$ (or $N_{R^{\prime}})$ as given by Eqn.(\ref{mass}) and (\ref{dm}). These are the operators in the augmented space $ \Omega = {\cal H} \otimes \Phi $. The state $|{R} \otimes \{\emptyset\}\rangle$ is 
a state in the  augmented space, which is the direct product of the real-space and the configuration-space bases.The configuration-space $ \Phi = \prod_{R}^{\otimes}\phi_{R} $ is of rank $2^{N}$ for a system of N-lattice sites with binary distribution. A basis in this space is denoted by the cardinality sequence $ \{{\cal C}\} = \{R_{1},R_{2},\ldots,R_{c}\} $  which gives us the positions where we have a $\vert\!\downarrow \rangle$ configuration. The configuration $\{\emptyset\}$ refers to a null cardinality sequence i.e. one in which we have $\vert\uparrow \rangle$ at all sites.

The {\sl virtual crystal } (VCA) Green matrix is :
\begin{equation}
 {\mbf g} ({R,R'},w^{2}) = \langle  \{\emptyset\}\otimes R \vert\ ( \ll \widetilde{\mbf M}\gg \omega^{2}  - \ll\widetilde{\bf D}\gg)^{-1}\ \vert \{\emptyset\} \otimes R'\rangle
\label{vcgreen}
\end{equation} 
\n where
\[ \ll\widetilde{\mbf M}\gg = \ll m\gg \widetilde{I}\otimes I \]
\noindent Referring back to Equations (\ref{mass}),(\ref{dm}) and (\ref{g0}) we get :

\begin{widetext}
\begin{eqnarray}
\ll {\mbf G}(R,R',w^2)\gg &\ =\ & 
 \langle  \{\emptyset\}\otimes R \vert\ \left( \rule{0mm}{4mm} \ll \widetilde{\mbf M}\gg \omega^{2}  - \ll\widetilde{\bf D}\gg +\widetilde{\mbf M}'\omega^2 - \widetilde{\mbf D}'\right)^{-1}\ \vert \{\emptyset\} \otimes R'\rangle \nonumber\\
& \ =\ &  \langle\{\emptyset\}\otimes R\vert \left({\mbf g}^{-1}- \widetilde{\mbf D}_1\right)^{-1}\vert
\{\emptyset\}\otimes R'\rangle 
\label{main}
\end{eqnarray}
\n we define :
\begin{eqnarray*}
\widetilde{\mathbf D}_1 = \sum_R \  \left\{\rule{0mm}{5mm} {- \mbf \Upsilon}_R -\sum_{R'\ne R} {\mathbf\Psi}_{RR'}\right\}\ \otimes P_R\ +\ 
\sum_{R}\sum_{R'\ne R}\  {\mathbf\Psi}_{RR'}\otimes T_{RR'}
\end{eqnarray*}
\n with
\begin{eqnarray}
{\mbf\Upsilon}_R & = & \B({\mbf m})\ w^2\ p^\downarrow_R \ +\ \F({\mbf m})\ w^2\ {\cal T}^{\uparrow\downarrow}_R
\nonumber\\
\phantom{x}\nonumber\\
 {\mathbf\Psi}_{RR'}& = & {\mathbf D}^{(1)}_{RR'}\  \left(p^\downarrow_R +  p^\downarrow_{R'}\right)
+ {\mathbf D}^{(2)}_{RR'}\ \left({\cal T}^{\uparrow\downarrow}_R + 
 {\cal T}^{\uparrow\downarrow}_{R'}\right) + {\mathbf D}^{(3)}_{RR'}\  p^\downarrow_R\  p^\downarrow_{R'}+\ldots  \nonumber\\
\phantom{x}\nonumber\\
 \phantom{xxxx}  
& & \ldots+ {\mathbf D}^{(4)}_{RR'}\ \left( p^\downarrow_R\ {\cal T}^{\uparrow\downarrow}_{R'} + 
 {\cal T}^{\uparrow\downarrow}_{R}\ p^\downarrow_{R'}\right) +  
 {\mathbf D}^{(5)}_{RR'}\ {\cal T}^{\uparrow\downarrow}_R
\ {\cal T}^{\uparrow\downarrow}_{R'}
\label{dyn}
\end{eqnarray}
\end{widetext}
\n where
\begin{eqnarray*}
{\mathbf D}^{(1)}&=&(y-x)\ \Phi_{(1)}^{\alpha\beta}\\
{\mathbf D}^{(2)}&=&\sqrt{xy}\ \Phi_{(1)}^{\alpha\beta}\\
{\mathbf D}^{(3)}&=&{(y-x)^{2}}\ \Phi_{(2)}^{\alpha\beta}\\
{\mathbf D}^{(4)}&=&\sqrt{xy}\ (y-x)\ \Phi_{(2)}^{\alpha\beta}\\
{\mathbf D}^{(5)}&=& xy\ \Phi_{(2)}^{\alpha\beta}\\
\end{eqnarray*}
%\end{widetext}

\subsection{Generalized Yonezawa-Matsubara scattering diagrams for the averaged Green function }

\begin{figure*}[t]
\includegraphics[width=13cm,height=8cm]{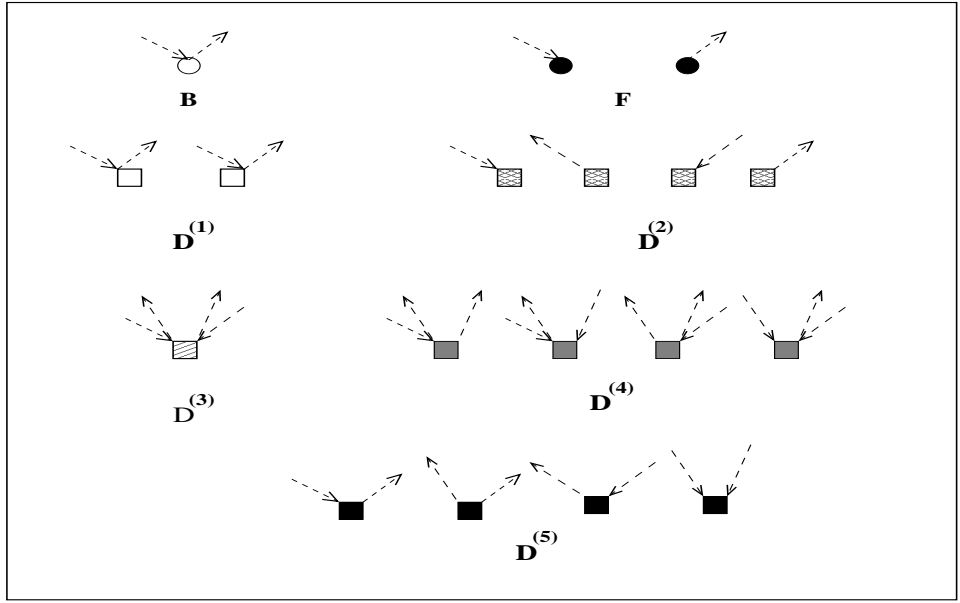}
\caption{The scattering vertices for the averaged Green function }
\label{fig1}
\end{figure*}
In this section we shall start from Eqn. (\ref{main}) and develop a multiple scattering picture based on this. The idea is very similar to that of Yonezawa and Matsubara \cite{ym} in the context of purely diagonal disorder. We shall first expand Eqn.(\ref{main}) as follows :
\begin{widetext}
\begin{eqnarray}
 \ll {\mbf G}(R,R',w^2)\gg 
 =  \langle \{\emptyset\}\otimes R \vert \left(\rule{0mm}{5mm} \g + \g\ \widetilde{\mbf D}_1\ \g 
+ \g\ \widetilde{\mbf D}_1\ \g\ \widetilde{\mbf D}_1\ \g + \ldots \right) \vert \{\emptyset\}\otimes R'\rangle \nonumber\\
\label{series}
\end{eqnarray}
\end{widetext}

Let us discuss very briefly how one generates the scattering diagrams. The first term in Equation (\ref{series}) gives :

\be
\langle \{\emptyset\}\otimes R \vert\ \g\  \vert  \{\emptyset\}\otimes R' \rangle \ =\ {\mbf g} (R,R',w^{2}) 
\label{ym1}
\ee

\n The second term yields zero since $\langle \{\emptyset\}\otimes R \vert \widetilde{\mbf D}_1 \vert \{\emptyset\}\otimes R' \rangle = 0 $. The third term gives~:

\begin{widetext}
\begin{eqnarray*}
 \sum_{S'S''}\sum_{S''' S''''}\sum_{\{\cal C\}}\sum_{\{\cal C'\}}\ \langle \{\emptyset\}\otimes R\vert\ \g\ \vert \{\emptyset\}\otimes S'\rangle\ \langle \{\emptyset\}\otimes S'\vert\ 
\wt{\mbf D}_1\ \vert \{{\cal C}\}\otimes S^{\prime\prime}\rangle\ \ldots\\
\ldots \langle \{{\cal C}\}\otimes S^{\prime\prime}\vert
\ \g\  \vert \{{\cal C}'\}\otimes S^{\prime\prime\prime}\rangle\ \langle \{{\cal C}'\}\otimes S^{\prime\prime\prime}\vert
\ \wt{\mbf D}_1\ \vert \{\emptyset\}\otimes S'''' \rangle\ \langle \{\emptyset\}\otimes S''''\vert\ \g\ \vert \{\emptyset\}\otimes R'\rangle 
\end{eqnarray*}
\mbox {A little algebra yields the following contribution :}
\begin{eqnarray}
 \sum_{S_1S_2}\ \g(R,S_1,w^2)\  (\F w^2)\  \g(S_1,S_2,w^2)\ \delta(S_1-S_2)\ (\F w^2)\ \g(S_2,R',w^2) + \ldots \nonumber\\
 \sum_{S_1S_2}\sum_{S_3S_4}\ \g(R,S_1,w^2)\  \D^{(2)}_{S_1S_2}\ \g(S_2,S_3,w^2)\ \delta(S_1-S_3)\ \D^{(2)}_{S_3S_4}\ \g(S_4,R',w^2) +\ldots \nonumber\\
 \sum_{S_1S_2}\sum_{S_3S_4}\ \g(R,S_1,w^2)\  \D^{(2)}_{S_1S_2}\ \g(S_2,S_3,w^2)\ \delta(S_2-S_3)\ \D^{(2)}_{S_3S_4}\ \g(S_4,R',w^2) +\ldots \nonumber\\
 \sum_{S_1S_2}\sum_{S_3S_4}\ \g(R,S_1,w^2)\  \D^{(2)}_{S_1S_2}\ \g(S_2,S_3,w^2)\ \delta(S_1-S_4)\ \D^{(2)}_{S_3S_4}\ \g(S_4,R',w^2) +\ldots \nonumber\\
 \sum_{S_1S_2}\sum_{S_3S_4}\ \g(R,S_1,w^2)\  \D^{(2)}_{S_1S_2}\ \g(S_2,S_3,w^2)\ \delta(S_2-S_4)\ \D^{(2)}_{S_3S_4}\ \g(S_4,R',w^2) +\ldots \nonumber\\
 \sum_{S_1S_2}\sum_{S_3S_4}\ \g(R,S_1,w^2)\  \D^{(5)}_{S_1S_2}\  \g(S_2,S_3,w^2)\ \delta(S_2-S_4)\delta(S_1-S_3)\ \D^{(5)}_{S_3S_4}\ \g(S_4,R',w^2) +\ldots \nonumber\\
 \sum_{S_1S_2}\sum_{S_3S_4}\ \g(R,S_1,w^2)\ \D^{(5)}_{S_1S_2}\ \g(S_2,S_3,w^2)\ \delta(S_1-S_4)\delta(S_2-S_3)\ \D^{(5)}_{S_3S_4}\ \g(S_4,R',w^2) +\ldots \nonumber\\
\label{ym3}
\end{eqnarray}
\end{widetext}
\n Referring to Eqns. (\ref{ym1})and (\ref{ym3}) we shall now build up the Yonezawa-Matsubara type diagrams. First
we shall associate scattering vertices with the terms in $\widetilde{\bf M}'$ and $\widetilde{\bf D}'$.
The Fig. \ref{fig1} shows the seven different type of scattering vertices. The dashed lines
are associated with the delta functions. 

\begin{figure}
\includegraphics[height=7cm,width=8cm]{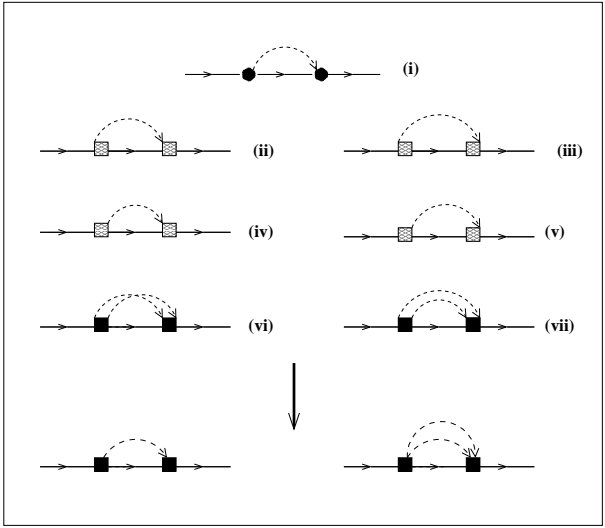}
\caption{The  scattering diagrams for n=2. The top four lines [ fig (i) to fig (vii) ] show all possible diagrams. The bottom line is a schematic representation of the the topologically distinct classes of diagrams}
\label{fig2}
\end{figure}

\n With each factor {\bf g} we shall associate a propagator represented by a horizontal arrow. The
connected diagrams to order $n$ are then built up by stringing together $n+1$ propagators connected
by $n$ vertices with all dashed {\sl fluctuation lines} connected in pairs. 
The algebraic terms in Eqn. (\ref {ym3}) are then represented by the diagram shown in Fig.\ref{fig2}.
The Yonezawa-Matsubara diagrams were originally developed only for diagonal disorder. The
diagrams shown in Fig. \ref{fig2} involve off-diagonal scattering terms and the associated diagrams
are generalized Yonezawa-Matsubara diagrams. 

\begin{figure}[t]
\centering
\includegraphics[height=6.5cm,width=8cm]{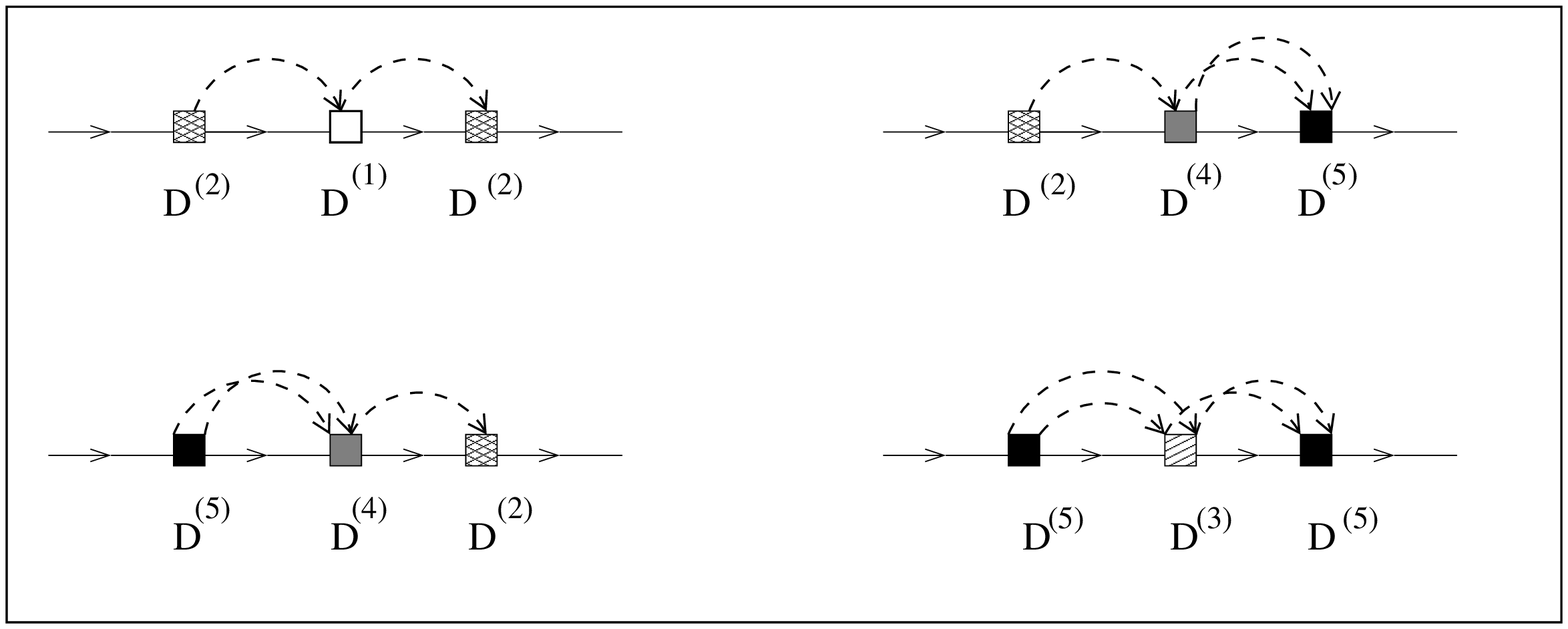}
\caption{The topologically distinct scattering diagrams for n=3.} 
\label{fig3}
\end{figure}

 Fig. \ref{fig3} shows the topologically distinct classes of diagrams for n=3. Note that it involves terms with contributions from {\bf D}$^{(1)}$, {\bf D}$^{(3)}$ and {\bf D}$^{(4)}$ . These scattering vertices cannot sit either in the leftmost or  in the rightmost positions, because one of the associated pseudo-fermion Green function line vanishes.

\begin{figure}[b]
\includegraphics[height=7cm,width=8cm]{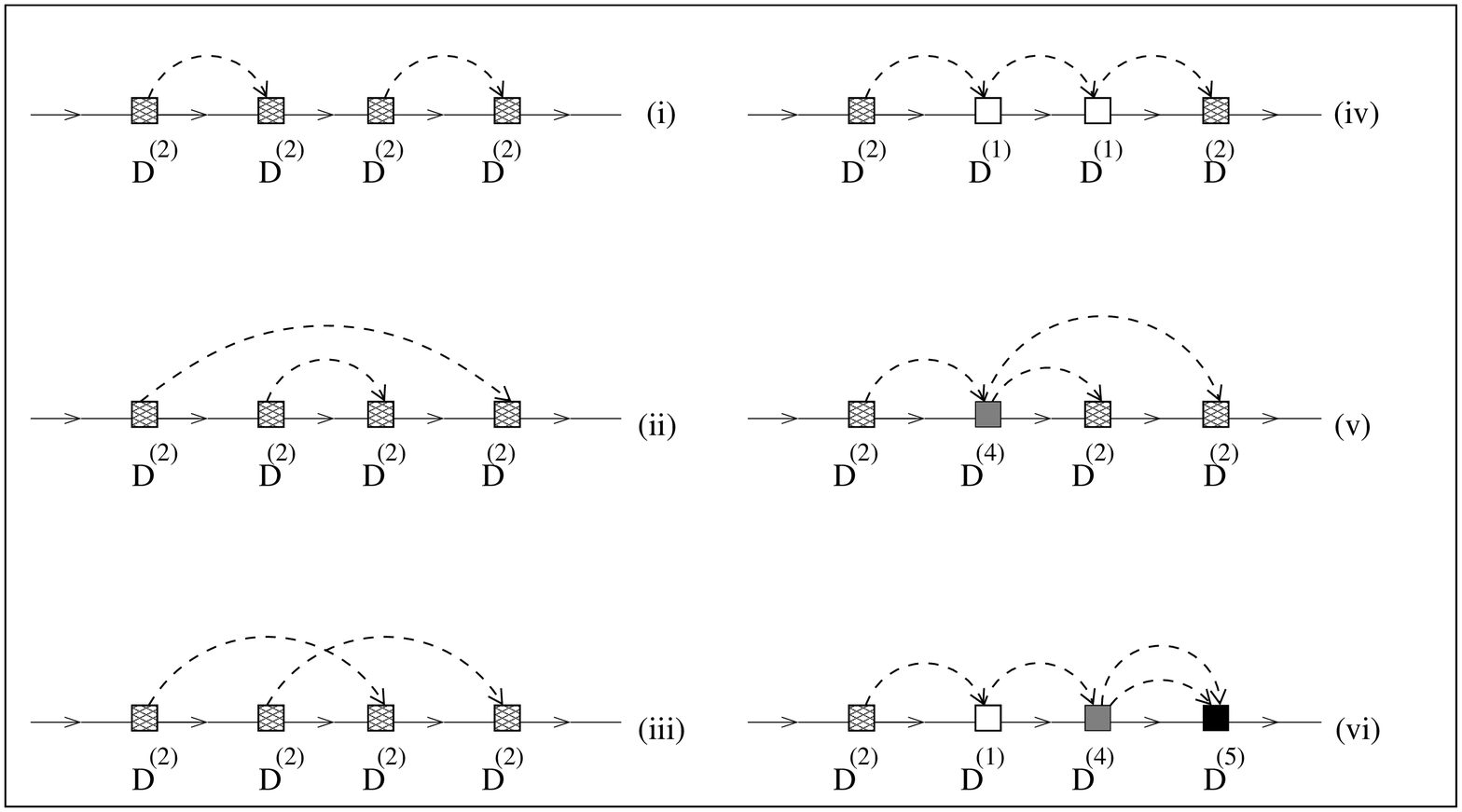}
\caption{The topologically distinct scattering diagrams for n=4. (i) A separable diagram involving
only {\bf D}$^{(2)}$ vertices. (ii) A non-separable, non-skeleton diagram involving {\bf D}$^{(2)}$ vertices. (iii)
A skeleton diagram involving {\bf D}$^{(2)}$ vertices. (iv)-(vi) Skeleton diagrams involving various combination
of vertices.}
\label{fig4}
\end{figure}

 For n=4, there are various classes of diagrams  involving {\bf D}$^{(2)}$
vertices as shown in Fig. \ref{fig4}. In this figure, (i) shows a separable double tent diagram \footnote{a separable
diagram is one that can be broken into two along a electron line without also breaking a pseudo-fermion line}. The second
tent goes to renormalize the rightmost phonon Green function from $\g(x,y)$ to $\ll {\bf G}(x,y\gg$. (ii) shows a double tent non-separable diagram
and (iii) a non-separable crossed-tent diagram. 
 Of these, the inner tent in the double tent  diagram (ii) goes on to renormalize the interior
Green function. As such, the crossed tent  diagram (iii) is a {\sl skeleton}
diagram of this class. 
The diagrams (iv)-(vi) are a few more examples of skeleton diagrams in this order involving other types of vertices.

If we club together the contribution of {\sl all} the skeleton diagrams calling  this the
self-energy,  and allow {\sl all} phonon Green functions except
the left-most to be renormalized by the separable and non-separable, non-skeleton diagrams, we get the Dyson equation~:
\[ \ll {\mbf G}\gg \eq \g +    \g\ {\mathbf \Sigma} \ll {\mbf G}\gg \]
For homogeneous disorder we have shown earlier that we have translational symmetry in the full augmented
space \cite{gdma}. We can then take Fourier transform of the above equation to get~:

\[  \ll\mbf{G}(\q,E)\gg \eq \mbf{g}(\q,E) + \mbf{g}(\q,E)\ \mbf{\Sigma}(\q,E)\ \ll\mbf{G}(\q,E)\gg
\label{dys1}\]

\begin{figure}[t]
\centering
\includegraphics[height=8cm,width=8cm]{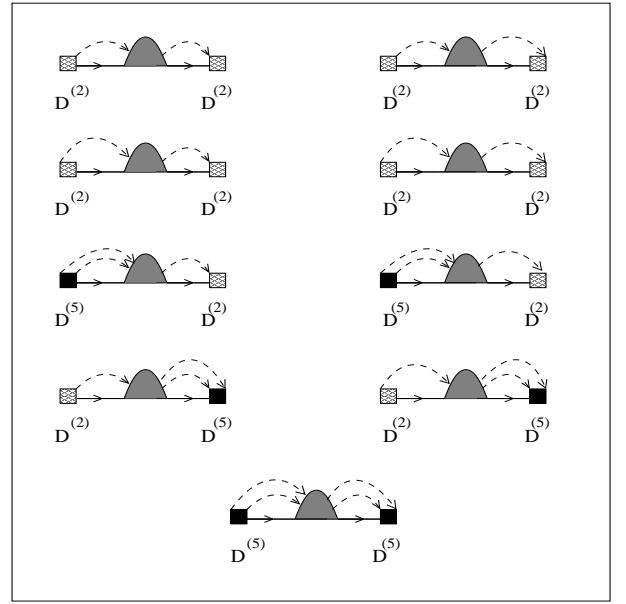}
\caption{Structure of the skeleton diagrams for the self-energy}
\label{fig5}
\end{figure}

The diagrams for the self-energy are skeleton diagrams {\sl all} of which have the structure as shown in Fig. \ref{fig5}.

\section{IMPORTANT PHYSICAL QUANTITIES}
In this section we shall derive results for important physical quantities such as the effective scattering length, {\it inelastic coherent and incoherent scattering cross sections.} All of these are derived based on the arguments of Nowak and Dederichs for the various kinds of scattering diagrams and it turns out that the numerical evaluation of these quantities require the entire configuration averaged Green matrices $\ll {\mbf G} ({\bf q},w)\gg$ and self energy matrices $\Sigma ({\bf q},w)$ in the reciprocal space representation.
\subsection{The inelastic neutron scattering cross-section}
The formal expression for the inelastic cross-section for the scattering of thermal neutrons from an initial state
labelled by {\bf k} to a final state {\bf k}$^\prime$ with a change of energy, 

\[ E = \hbar w = \frac{\hbar^2}{2M_n} (k^2-k'^2)\]

\n and a change of wave-vector ${\bf q} = {\bf k}-{\bf k'}+{\bf Q}$ , where {\bf Q} is a reciprocal lattice vector is~:

\begin{widetext}
\begin{eqnarray}
 \frac{d^2\sigma}{d\Omega dE}\ =\ \frac{1}{2N\hbar} \frac{k'}{k} \sum_R\sum_{R'}\sum_{\alpha\beta}\  q^\alpha q^\beta\ 
\left(\rule{0mm}{4mm} W_R\ \Im m\ G_{RR'}^{\alpha\beta}(w)\ W_{R'}\right)\  n(w)\ \exp{\{i{\mathbf q}\cdot(R-R')\}}\nonumber\\
\end{eqnarray}
\end{widetext}

\n here : $W_R = w_R\left\{\exp[-(1/2)\langle({\bf q}\cdot{\bf u}_R)^2\rangle_{th}]\right\}$ : $w_R$ is the scattering length of the nucleus of the atom sitting at $R$, its equilibrium position,  and {\bf u}$_R$(t) is its deviation from equilibrium at the time $t$. $n(w)$ is the Bose distribution function. For a random alloy, $w_R$, the Debye-Waller factor, the atomic mass and the force constants are all random variables and dependent on one another via the random occupation variables $\{n_R\}$. Carrying out averaging over nuclear spins as well as over all the random configurations~:

\begin{widetext}
\begin{eqnarray}
 \left[\frac{d^2\sigma}{d\Omega dE}\right]_{av}\ =\ \frac{1}{2N\hbar} \frac{k'}{k} \sum_R\sum_{R'}\sum_{\alpha\beta}\  q^\alpha q^\beta\ 
\Im m\ \left[\rule{0mm}{4mm} W_R\  G_{RR'}^{\alpha\beta}(w)\ W_{R'}\right]_{av}\  n(w)\ \exp{\{i{\mathbf q}\cdot(R-R')\}}\nonumber\\
\end{eqnarray}
\end{widetext}

Given homogeneity of disorder, we may rewrite the above configuration-average as~:

\begin{equation}
 \left[\frac{d^2\sigma}{d\Omega dE}\right]_{av} = \frac{1}{2\hbar} \frac{k'}{k} \sum_{\alpha\beta}\  q^\alpha q^\beta\ \Im m\ {\cal G}^{\alpha\beta}({\mathbf q},w)\  n(w)\ \nonumber\\
\end{equation}
\begin{eqnarray}
\left[W_R\ G^{\alpha\beta}_{RR'}(w)\ W_{R'}\right]_{av} =& \hspace{-0.75in} {\cal G}^{\alpha\beta}(R-R',w) \nonumber\\
{\cal G}^{\alpha\beta}({\mathbf q},w) =& \sum_{R}\ {\cal G}^{\alpha\beta}(R,w)\ \exp{\{ i{\mathbf q}\cdot R\}}\nonumber\\
\end{eqnarray}

\begin{figure}[b]
\centering
\includegraphics[height=3cm,width=5cm]{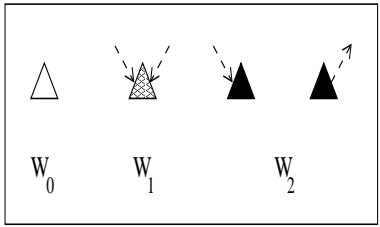}
\caption{The scattering vertices related to the fluctuations in $W$}
\label{fig6}
\end{figure}

Since $W_{R}$ is a random variable taking two values $W_{A}$ and $W_{B}$ depending on which kind of atom sits at the site labeled 'R', So we can write $W_R = W_A\ n_R + W_B\ (1-n_R)$. Augmented space theorem then leads to  the $W$ factor being replaced by an operator in configuration-space as:

\begin{eqnarray} \widetilde{\mbf W}& =&  \A(W) \tilde{I}  + {\mbf B(W)} \sum_R P_R\otimes p^\downarrow_R  + \F(W) \sum_R P_R\otimes
{\cal T}^{\uparrow\downarrow}_R \nonumber\\
 & =&  {\mbf W}_0 \tilde{I}  + {\mbf W}_1 \sum_R P_R\otimes p^\downarrow_R  + {\mathbf W}_2 \sum_R P_R\otimes
{\cal T}^{\uparrow\downarrow}_R \nonumber\\
\label{w1}\end{eqnarray}

\n Since, like the mass, $W$ is a scalar, mode independent quantity, {\bf A}(W) = $A(W)\ \delta_{\alpha\beta}$, 
 {\bf B}(W) = $B(W)\ \delta_{\alpha\beta}$, 
 {\bf F}(W) = $F(W)\ \delta_{\alpha\beta}$ and $\widetilde{\bf W}$ = $\widetilde{W}\ \delta_{\alpha\beta}$.

\n The scattering vertices arising out of Eqn.(\ref{w1})  are shown in Fig.\ref{fig6}.
%\vskip 0.2cm
\begin{widetext}
\n The augmented space theorem then gives : 
\begin{eqnarray}   \ll {\mbf W}_R {\mathbf G}_{RR'} {\mathbf W}_{R'}\gg
  =\langle R\otimes \{\emptyset\}\vert 
\widetilde{\mbf W}\ \left(\rule{0mm}{4mm}{\mbf g} + {\mbf g} \widetilde{\mbf D}_1 {\mbf g} + {\mbf g} \widetilde{\mbf D}_1  {\mbf g}
\ \widetilde{\mbf D}_1\ {\mbf g}+\ldots \right)\widetilde{\mbf W}\vert R'\otimes \{\emptyset\}\rangle \nonumber\\
\label{fd}
\end{eqnarray}
\end{widetext}

\begin{figure*}
\centering
\includegraphics[height=13cm,width=15cm]{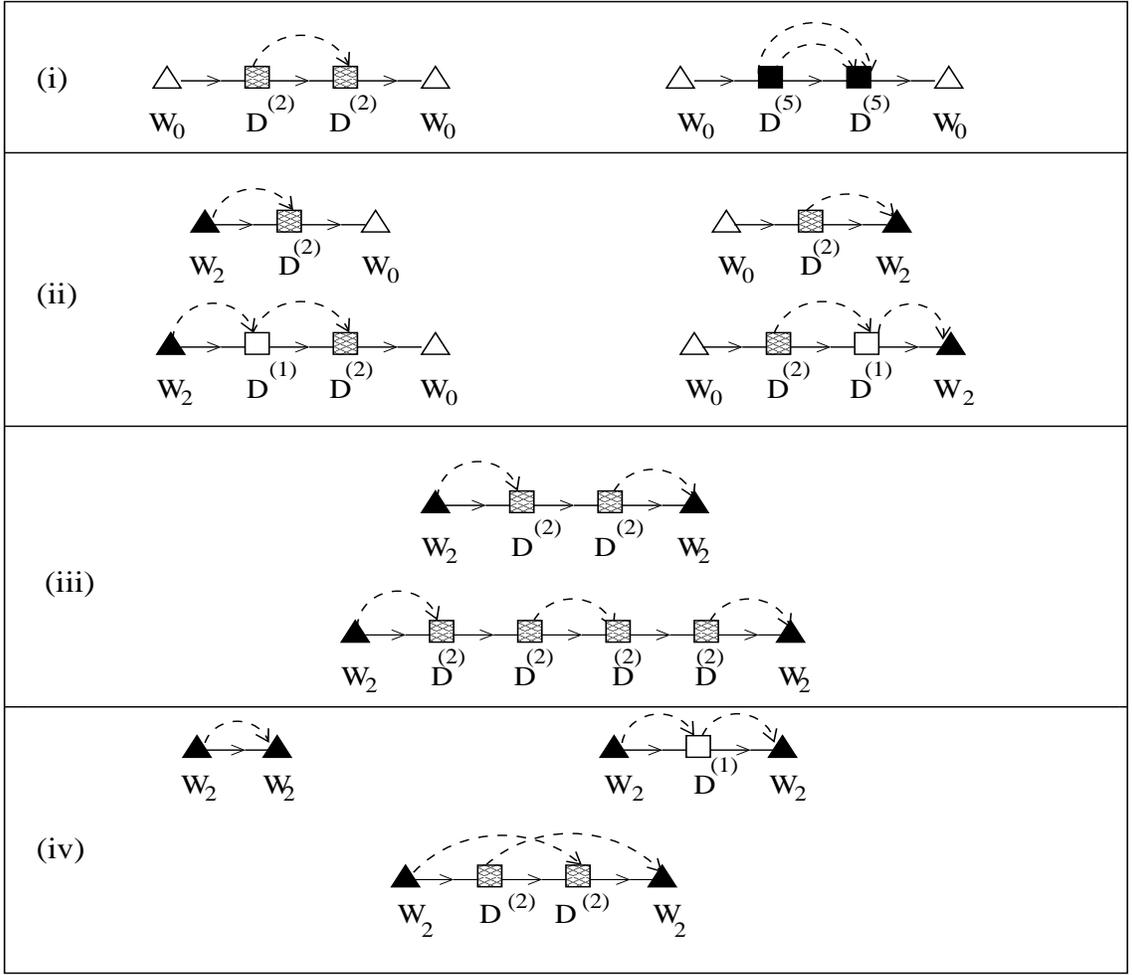}
\caption{The scattering diagrams for the inelastic scattering cross-section}
\label{fig7}
\end{figure*}

\n The Fig. \ref{fig7} shows the scattering diagrams produced from Eqn.(\ref{fd}) for the scattering cross-section. We have grouped them
into four categories :
\begin{enumerate}
\item In the category (i) are reducible diagrams whose end vertices are $\W_0$ or the averaged value  $\ll W\gg\ \delta_{\alpha\beta} $. The central decorations are all the diagrams we have already seen for the configuration-averaged Green function. The contribution of these diagrams
are therefore :

\[ \W_0\ \ll {\mathbf G}(w)\gg \W_0 \]

\item The second set of diagrams (ii) are also reducible diagrams. Inspection of the diagrams immediately show us that their
contribution may be written as~:

\begin{eqnarray*}
\W'(w) \ll {\mbf G}(w)\gg  \W_0 +\ldots\\
\ldots   \W_0\ \ll {\mbf G}(w)\gg\ \W^{\prime\prime}(w) \end{eqnarray*}

The contribution of the 'vertex' $\W'(w)$ and $\W^{\prime\prime}(w)$ are  closely related to the self-energy, with the initial and final
vertices being different : $\W_2$ rather than {\bf D}$_2$ or {\bf D}$_5$. We shall discuss this relationship subsequently.

\item The third set of diagrams (iii) are reducible diagrams with contribution :

\[ \W'(w)\ \ll {\mbf G}(w)\gg\ \W^{\prime\prime}(w) \]

The contribution of these three sets of reducible diagrams may be added together to give :

\[
 \W_{eff}(w)\ \ll {\mbf G}(w)\gg\ \widehat{\W}_{eff}(w)
\]
\n where

\[  \W_{eff}(w)=\W_0+\W'(w)\]
\n and 
\[\widehat{\W}_{eff}(w)=\W_0+\W^{\prime\prime}(w)
\]

\item The last class of diagrams (iv) are irreducible diagrams. Their contribution is also related to the self-energy
with both the initial and final vertices being $\W_2$ rather than {\bf D}$^{(2)}$ or {\bf D}$^{(5)}$. We shall discuss these diagrams
in detail subsequently.
\end{enumerate}

\n Because of the disorder renormalization, the $\W_{eff}$ is diagonal neither in real nor mode space and becomes frequency
dependent and complex.
The reducible diagrams contributes to an expression :

\begin{equation}
{\cal G}^{\alpha\beta}_{red}({\mbf q},w)\ =\ \sum_{\mu\nu}\ W_{eff}^{\alpha\mu}({\mbf q},w)\ \ll G^{\mu\nu}({\mbf q},w)\gg\ 
\widehat{W}_{eff}^{\nu\beta}({\mbf q},w)
\label{coh_red}
\end{equation}

\n If we now examine the structure of the self-energy diagrams in Fig. \ref{fig5}, we note that the vertex {\bf D}$_2$
creates one configuration fluctuation at a site if it is an initial vertex and annihilates a configuration fluctuation if it is
a final vertex. On the other hand, the {\bf D}$_5$ vertex creates {\sl two} configuration fluctuations at two sites 
if it is an initial vertex and annihilates {\sl two} configuration fluctuations at two sites if it is a final vertex. The
diagrams for both $\W_{eff}(w)$ and the irreducible diagrams in Fig. \ref{fig7} have vertices which create or annihilate only
one configuration fluctuation at both the initial and final vertices. If we denote the part of the self-energy contributed
by the diagrams in the first two lines of Fig. \ref{fig5} by ${\bf \Sigma}(w)$, then it follows that :

\begin{eqnarray}
{\mbf W}_{eff}({\mbf q},w)\ =\ {\mbf W}_0\ +\ {\mbf W}_2\ [{\mbf \Delta}({\mbf q})]^{-1}\ {\mbf \Sigma}({\mbf q},w)\nonumber\\
\widehat{\mbf W}_{eff}({\mbf q},w)\ =\ {\mbf W}_0\ +\ {\mbf \Sigma}({\mbf q},w)  [{\mbf \Delta}({\mbf q})]^{-1}\ {\mbf W}_2
\label{length}
\end{eqnarray}
where
\begin{equation}
{\mbf \Delta }({\bf q}) = {\mbf F}({\mbf m}) + {\mbf D}^{(2)}({\bf q}) + {\mbf D}^{(5)}({\bf q})\nonumber
\end{equation}

The expression (\ref{coh_red}) is long-ranged in real-space. Following the argument of Nowak and Dederichs \cite{nd} within
the CPA and Mookerjee and Yussouff \cite{my} in a more general context of a cluster-CPA, we identify contribution of
the reducible part as the coherent part of the inelastic scattering :
\begin{widetext}
\begin{eqnarray}
 \left[\frac{d^2\sigma}{d\Omega dE}\right]_{av}^{coh}\ =\ 
 \frac{1}{2\hbar} \frac{k'}{k} \sum_{\alpha\beta}\  q^\alpha q^\beta\ 
\Im m\ \left[\rule{0mm}{4mm} {\mbf W}_{eff}({\mbf q},w) \ll {\mbf G}({\mbf q},w)\gg\ \widehat{\mbf W}_{eff}({\mbf q},w)\right]^{\alpha\beta}\  n(w)\nonumber\\
\label{coh}
\end{eqnarray}
\end{widetext}

\n If we now look back at the irreducible diagrams in class (iv) of Fig. \ref{fig7}, we note that the diagrams in the top row of (iv) are both short ranged : the leftmost one is totally diagonal in real-space and the rightmost one has the
same range as the dynamical matrices, which are reasonably short ranged. The bottom diagram is longer ranged. However, 
this and {\sl all} long ranged diagrams in this class involve crossed-tent diagrams. If we look at the diagrams for
the self-energy exactly the same kind of structure is seen. The only diagrams which can lead to a long-ranged self-energy
are crossed-tent diagrams like (iii) of Fig. \ref{fig4}.  These diagrams belong to correlated
scattering from configuration fluctuations at different sites. The contributions of such diagrams are dominated by
those which are short-ranged. Within the single site coherent potential approximation (CPA) such diagrams are neglected altogether and the self-energy (and therefore the irreducible diagrams for the cross-section) is diagonal in real-space. 
 Beyond the CPA, dominant contributions arise from correlated scattering of the smaller clusters and the contribution of the irreducible diagrams is also short-ranged : being between sites within the smaller clusters. The range of the self-energy and the irreducible diagrams are therefore as large as the size of the largest cluster whose correlated scattering is significant. The {\sl Locality principle} of Heine \cite{heine} convinces us that electronic structure is insensitive of far-off environment and, although the self-energy is not diagonal in real-space (except in the single-site CP approximation), its range is nevertheless short. This is behind the reasonable success of the
CPA in any cases. Again, following the arguments of Nowak and Dederichs \cite{nd} and Mookerjee and Yussouff \cite{my}, this
contribution can be related to the incoherent part of the inelastic scattering. The incoherent intensity is given by :
\begin{eqnarray}
 \left[\frac{d^2\sigma}{d\Omega dE}\right]_{av}^{incoh}\ =\ 
 \frac{1}{2\hbar} \frac{k'}{k} \sum_{\alpha\beta}\  q^\alpha q^\beta\ 
\Im m\ \left[\rule{0mm}{4mm} {\mbf \Gamma}({\mbf q},w)\right]^{\alpha\beta}\  n(w)\nonumber\\
\label{incoh}
\end{eqnarray}

\n where 

\begin{equation}
{\mbf \Gamma}({\mbf q},w) \ =\ \W_2\ [{\mbf\Delta}({\mbf q})]^{-1}\ {\mbf \Sigma}({\mbf q},w) \ [{\mbf \Delta}({\mbf q})]^{-1}\ \W_2
\end{equation}

\section{Block recursion and the self-energy matrix}

The augmented space recursion (ASR)  has been proposed earlier
by us \cite{kasr, alam} as  technique for the incorporation of the effects of
configuration fluctuations 
 for random substitutionally disordered alloys.
 This can be achieved without the usual problems of violation
of the herglotz analytic properties \cite{alam} of the approximated configuration averaged Green functions
for phonon problems in realistic random alloys. Although our initial focus was
on  spectral functions and complex dispersion relations and lifetimes, in this
communication we propose to study 
inealstic neutron scattering cross-sections. These
calculations require the full Green matrices  and not only their
diagonal elements. We propose here the use of a generalization of the recursion method of Haydock {\sl et.al.} \cite{rec}. The block recursion technique had been introduced earlier by Godin and Haydock \cite{godin1,godin2} in the very different context for obtaining the scattering S-matrix for finite scatterers attached to perfect leads. We shall borrow their ideas and set up a
block recursion in the space of vibrational modes (rather than the lead space, as in Godin and Haydock's work) in order to obtain the Green matrices  directly.

The recursion method essentially starts from a denumerably infinite basis and
changes the basis to one in which the dynamical matrix (or the Hamiltonian, in
electronic problems) is tri-diagonal. In the block recursion we start from
a matrix basis of the form : $\{\Phi^{(n)}_{J,\alpha\beta}\}$, where $J$ is the discrete labelling of the augmented space states and the $\alpha,\beta$ labels
Cartesian directions (i.e. the modes of vibration).
The inner product of such basis is defined by :

\[\left( \Phi^{(n)}, \Phi^{(m)}\right) \ =\ \sum_J\sum_{\beta'}\ \Phi^{(n)\dagger}_{\alpha\beta',J}\ 
\Phi^{(m)}_{J,\beta'\beta}\ =\ N^{nm}_{\alpha\beta} \]

For a real-space calculations on a lattice with Z-nearest neighbours, we  start the recursion with :

\[ \Phi^{(1)}_{J,\alpha\beta}\ =\ U_{\alpha\beta}^{(1)} \ \delta_{J,1} + U_{\alpha\beta}^{(2)} \ \delta_{J,Z+1} \] 

while for a reciprocal space calculation we start with :

\[ \Phi^{(1)}_{J,\alpha\beta}\ =\ U_{\alpha\beta}^{(1)} \ \delta_{J,1} + U_{\alpha\beta}^{(2)} \ \delta_{J,2} \]

\n where

$U_{\alpha\beta}^{(1)}=\frac{A(m^{-1/2})}{\left[\rule{0mm}{2mm}A(m^{-1})\right]^{1/2}}\ \delta_{\alpha\beta}\ \ \ ;\ \ \ U_{\alpha\beta}^{(2)}=\frac{F(m^{-1/2})}{\left[\rule{0mm}{2mm}A(m^{-1})\right]^{1/2}}\ \delta_{\alpha\beta}$

The remaining terms in the basis are recursively obtained from :
\begin{widetext}
\begin{eqnarray}
 \sum_{\beta'}\Phi^{(2)}_{J,\alpha\beta'} B^{(2)\dagger}_{\beta'\beta}& = &\sum_{J'}\sum_{\beta'} \widetilde{H}_{J\alpha,J'\beta '} \Phi^{(1)}_{J',\beta'\beta} - \sum_{\beta'}\Phi^{(1)}_{J,\alpha\beta '} A^{(1)}_{\beta '\beta}\nonumber\\
\sum_{\beta '} \Phi^{(n+1)}_{J,\alpha\beta '} B^{(n+1)\dagger}_{\beta '\beta}& = &\sum_{J'}\sum_{\beta'} \widetilde{H}_{J\alpha,J'\beta '} \Phi^{(n)}_{J',\beta '\beta}- \sum_{\beta '}\Phi^{(n)}_{J,\alpha\beta '} A^{(n)}_{\beta '\beta} - \sum_{\beta '}\Phi^{(n-1)}_{J,\alpha\beta '} B^{(n)}_{\beta '\beta}\nonumber\\
\label{eqn24}
\end{eqnarray}
\end{widetext}
where,\ \ $ \widetilde{\mathbf H}\ =\ \widetilde{\mbf M}^{-1/2}\ \widetilde{\mbf D}\ \widetilde{\mbf M}^{-1/2} $  with $\widetilde{\mbf M}$ and $\widetilde{\mbf D}$ given by Eqns.(\ref{mass}) and (\ref{dm}).

\n Orthogonalization of the basis gives :

\begin{equation} 
  \sum_{J}\sum_{\beta '}\sum_{J'}\sum_{\beta ''}\Phi^{(n)\dagger}_{\alpha\beta ',J}\ \widetilde{H}_{J\beta', J'\beta ''} \Phi^{(n)}_{J',\beta''\beta '}=\sum_{\beta'}\ N^{nn}_{\alpha\beta'}\ A^{(n)}_{\beta '\beta}\nonumber
\end{equation}

\n In matrix notation, where matrices are in the vibrational mode ($\alpha\beta$) space :

\begin{equation}
{\mathbf A}^{(n)} \ =\ \left({\mathbf N}^{nn}\rule{0mm}{4mm}\right)^{-1}\ \sum_{J}\sum_{J'}\ {\mathbf \Phi}^{(n)\dagger}_J\ \widetilde{\mathbf H}_{JJ'}\ {\mathbf \Phi}^{(n)}_{J'} 
\end{equation}

Next, we note that we had started with a orthogonal basis set of rank $J_{max}\times \alpha_{max}$ .
 The above procedure merely gives $J_{max}$ basis sets. We still have orthogonality conditions among
the various columns of $\Phi^{(n)}_{J,\alpha\beta}$. In order to impose these conditions, consider 
\begin{eqnarray*}
 \Psi_{J,\alpha\beta} = \sum_{J'}\sum_{\beta '} \widetilde{H}_{J\alpha,J'\beta '} 
\Phi^{(n)}_{J',\beta '\beta} - \sum_{\beta '}\Phi^{(n)}_{J,\alpha\beta '}\ A^{(n)}_{\beta '\beta}-\ldots\\
 - \sum_{\beta '}\Phi^{(n-1)}_{J,\alpha\beta '}\
B^{(n)}_{\beta '\beta}\end{eqnarray*}

\n Construct three column vectors  $\psi_{J\alpha}^{(\beta)}$ out of the three columns of $\Psi_{J,\alpha\beta}$ and set about to Gram-Schimidt orthonormalizing the set :

\begin{widetext}
\begin{eqnarray}
\psi^{(1)}_{\alpha J}  &=& \phi^{(1)}_{\alpha J}\ B_{11} \quad\Rightarrow\quad
B_{11}^2 \ =\ \sum_{\alpha J}{\psi^{(1)}_{J\alpha}}^{*} \psi^{(1)}_{\alpha J}\nonumber\\
\psi^{(2)}_{\alpha J}  &=& \phi^{(1)}_{\alpha J}\ B_{12} + \phi^{(2)}_{\alpha J}\ B_{22} \quad\Rightarrow\quad B_{12} \ =\ \sum_{\alpha J} {\phi^{(1)}_{J\alpha }}^{*}\psi^{(2)}_{\alpha J} \quad ;\quad B_{22}^2  =  \sum_{\alpha J}{\psi^{(2)}_{J\alpha}}^{*} \psi^{(2)}_{\alpha J} - B_{12}^2\nonumber\\
& &\ldots\ldots\ldots\ldots\ldots\ldots\ldots\nonumber\\
& &\ldots\ldots\ldots\ldots\ldots\ldots\ldots\nonumber\\
& &\ldots\ldots\ldots\ldots\ldots\ldots\ldots\nonumber\\
\psi^{(m)}_{\alpha J}  &=& \sum_{k=1}^{m}\ \phi^{(k)}_{\alpha J} B_{km} \quad\Rightarrow\quad B_{km} = \sum_{\alpha J} {\phi^{(k)}_{J\alpha}}^{*}\psi^{(m)}_{\alpha J} \quad (k < m) \quad ;\quad B_{mm}^2  = \sum_{\alpha J} \psi^{(m)}_{J\alpha} \psi^{(m)}_{\alpha J} - \sum_{k=1}^{m-1} B_{km}^{2}\nonumber\\
\label{eqn26}
\end{eqnarray}
\end{widetext}
\n where m stands for the total number of vibrational modes.

We may now  construct  $\Phi^{(n+1)}_{J,\alpha\beta}$ out of $\phi_{J\alpha}^{\beta}$ and note that   $B_{km}$ is indeed the matrix ${{\bf B}^{(n+1)}}^{\dagger}$ we are looking for.

The Eqns.(\ref{eqn24})-(\ref{eqn26}) show that we may calculate the matrices $\{{\bf A}^{(n)}, {\bf B}^{(n+1)}\}$ recursively, noting that ${\bf B}^{(1)}={\bf I}$ and ${\bf B}^{(0)}={\bf 0}$.
In this new basis, the Hamiltonian is  {\sl block tri-diagonal} and the Green matrix can be written as follows :

\begin{eqnarray}
{\mathbf G}^{(n)}&=& \left[\rule{0mm}{4mm} w^2\ {\mathbf I}- {\mathbf A}^{(n)}- {\mathbf B}^{(n+1)\dagger}\ {\mathbf G}^{(n+1)}\ {\mathbf B}^{(n+1)}\right]^{-1} \nonumber\\
& \phantom{x}& \nonumber\\
\ll {\mathbf G}\gg  &=&  {\mathbf G}^{(1)}
\end{eqnarray}
\n The terminator which replaces the asymptotic part of the matrix continued fraction is
that which is used by Godin and Haydock \cite{godin2}. We calculate the matrix coefficients
upto a $n=N_0$ and approximate at coefficients $> N_0$ by {\bf A} and {\bf B}. We then
write  for a $N \gg N_0$ :

\[ {\mathbf G}^{(N)}\ =\ \left[\rule{0mm}{4mm} (w^2 - i\delta){\mathbf I}\right]^{-1} \]

\n and then iterate :

\begin{eqnarray*}
G^{(n)} = \left[\rule{0mm}{4mm} w^2{\mathbf I}\ -\ {\mathbf A}\ -\ {\mathbf B}^\dagger\ {\mathbf G}^{(n+1)}\ {\mathbf B}\right]^{-1}& \\
 &  \mbox{ for } n > N_0 
\end{eqnarray*}

A judicious choice of  $\delta$ (0.001) and $N$ (5000)  gives a smooth density of states from the diagonal
part of the Green matrix. The self-energy follows from the  Dyson equation :
\[ {\mathbf \Sigma}\ =\ {\mathbf g}^{-1}\ -\ {\mathbf G}^{-1} \]

\begin{figure}
\centering
\includegraphics[width=8.5cm,height=15cm]{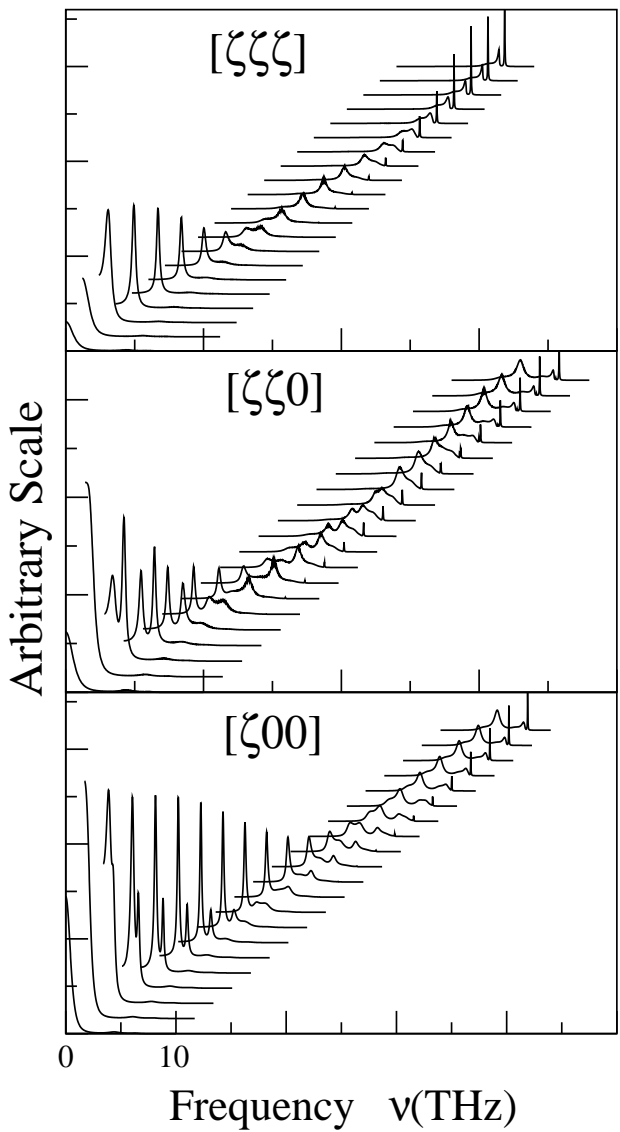}
\caption{The coherent scattering cross section in different directions for $Ni_{55}Pd_{45}$. In each of the different directions, the various curves indicate the cross sections for various $\zeta$ values starting from the lowest value to the edge of the Brillouin zone. The y-axis is in an arbitrary scale with heights scaled to the maximum height. Different curves for different $\zeta$ values are shifted along the x-axis in order to facilitate vision.}
\label{fig8}
\end{figure}

\section{Results and Discussion}
In the next two subsection, we explore the relative importance of mass and force constant disorder in the inelastic neutron scattering for two specific random alloys $Ni_{55}Pd_{45}$ and $Ni_{50}Pt_{50}$. In a recent publication \cite{alam}, we have already studied the advantages of Augmented space recursion (ASR) over the simple CPA for understanding the dispersion and life time of phonons in random binary alloys. The present work is an extension of that work from the implementation point of view, since now we need to apply the Block recursion technique to calculate the full Green matrix.
\subsection{${\bf Ni_{55}Pd_{45}}$ alloy : Strong mass and weak force constant disorder}

The mass disorder in this alloy is much larger than the force constant disorder. We refer the reader to a previous article \cite{alam} by us, for some of the basic properties of fcc Ni and Pd, which is relevant for our present calculation. This particular alloy has already been studied experimentally by Kamitakahara and Brockhouse \cite{kambrock} using Inelastic neutron scattering. The properties associated with the neutron scattering of Ni and Pd are reasonably favorable. The incoherent scattering cross section of Ni is fairly high, which was at first thought to be a potentially serious disadvantage, but in practice, this did not turn out to be much of a problem. it has been found that the scattered neutron distributions are always dominated by the coherent scattering even for high frequency made with large widths. Experimental investigation shows that significant difference between the coherent scattering lengths of Ni and Pd (The coherent scattering length for Ni is 1.03 while that of Pd is 0.6 in units of $10^{-12}$cm ) produces additional incoherent scattering in $Ni_{55}Pd_{45}$, but this is much smaller than the incoherence produced by the Ni-atom itself in the alloy.

In Fig.(\ref{fig8}), we display the inelastic coherent scattering cross sections [calculated from Eqn. (\ref{coh})] obtained from our Block recursion calculation along the highest symmetry directions $ ( [\zeta 0 0],[\zeta \zeta 0],[\zeta \zeta \zeta]\ ,\ \zeta = |\vec q|/|\vec {q}_{max}| ) $. For a particular direction, different curves indicate the cross sections at various $\zeta$-points starting from the lowest value ($\zeta=0$) to the edge of the Brillouin zone ($\zeta=1$ in units of 2$\pi$/a). The first thing to note is that the scattering cross sections are often asymmetric near the resonances. This property was also reflected in the phonon line shapes shown in the previous communication \cite{alam} by us. However the amount of asymmetries in the cross section is more than that in the usual lorentzian phonon line. That should be obvious because if we see the paper of Nowak and Dederichs \cite{nd} ; there they have derived an expression for the coherent scattering cross section in the single site CPA framework, this expression contains in addition to the usual lorentzian phonon line contribution [\ obtained from  $\Im m \ll G(\mbf {q},w)\gg$\ ], a second term which will have zero contribution only if the scattering length do not fluctuate (\ which is not the case in our formulation\ ). They have also argued that this extra term leads to an asymmetry (rather small contribution) of the phonon line. One can also notice that the nature of asymmetry in the cross section is not the same as in the line shapes. This is due to the contribution of off-diagonal elements of the Green's matrix and self energy matrix in the coherent scattering cross section [\ as obvious from expression (\ref{coh})\ ]. This difference in the nature of asymmetry is more pronounced in the $[\zeta \zeta 0]$ and $[\zeta \zeta \zeta]$ symmetry directions, because the Green's matrix [\ and self energy matrix\ ] comes out to be completely diagonal in the $[\zeta 0 0]$ direction. The occurence of such a structure of the cross section may also be due to the calculation in the mixed mode frame work. One should notice from the Block recursion technique described in section (IV) that, unlike the ordinary recursion where one extracts results for specific modes, the Block recursion requires a mixed mode starting state and hence does the calculation in that frame work to evaluate the entire Green matrix. The asymmetries can be described as a tendency of more scattering to occur near the resonance frequencies. It is important to note that the coherent scatteirng cross sections have a pronounced ${\mbf {q}}$-dependence in all the three symmetry directions. Because of the short range properties, the self energy (\ and $[\mbf {\Delta(q)}]^{-1}$\ )  depends only rather weakly on ${\mbf{q}}$ and does not show any strong structure as a function of the same. The same applies for the effective scattering length $\W_{eff}$. Thus the only strong ${\mbf q}$-dependence in the coherent cross section arises from the average Green's matrix $\ll \mbf {G(q,w)}\gg$ which is a long range matrix due to it's dependence on reducible diagrams.

\begin{figure}
\centering
\includegraphics[width=8.5cm,height=14cm]{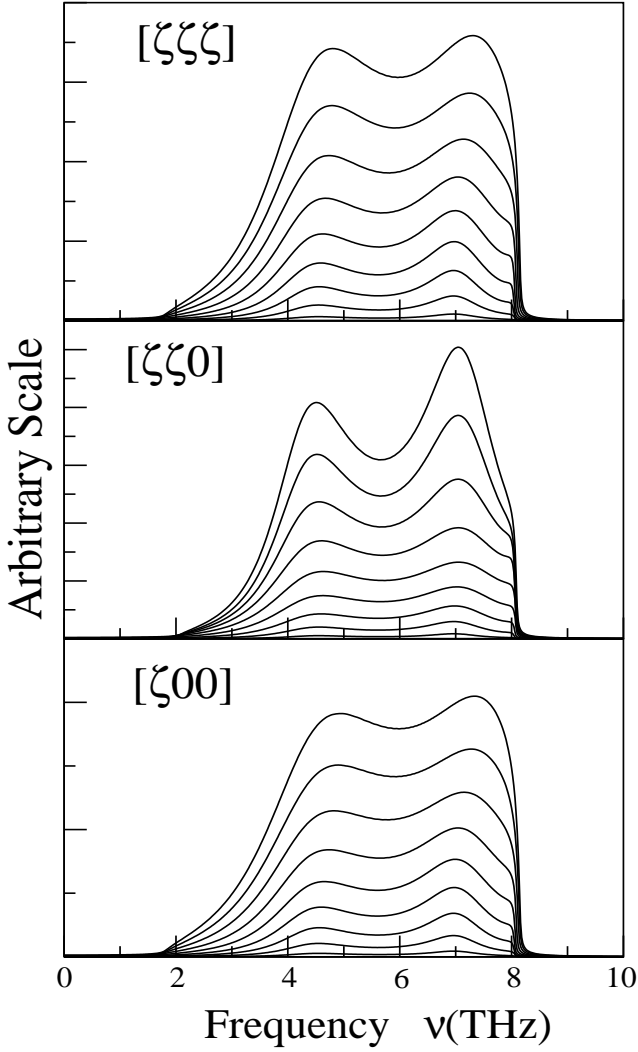}
\caption{The incoherent scattering cross sections in different directions for $Ni_{55}Pd_{45}$ alloy with $\Phi_{Ni-Pd}=0.7\ \Phi_{Ni-Ni}\ $. In each of the different directions, the various curves indicate the cross sections for various $\zeta$ values starting from the lowest value to the edge of the Brillouin zone.}
\label{fig9}
\end{figure}

\begin{figure}
\centering
\includegraphics[width=8.5cm,height=14cm]{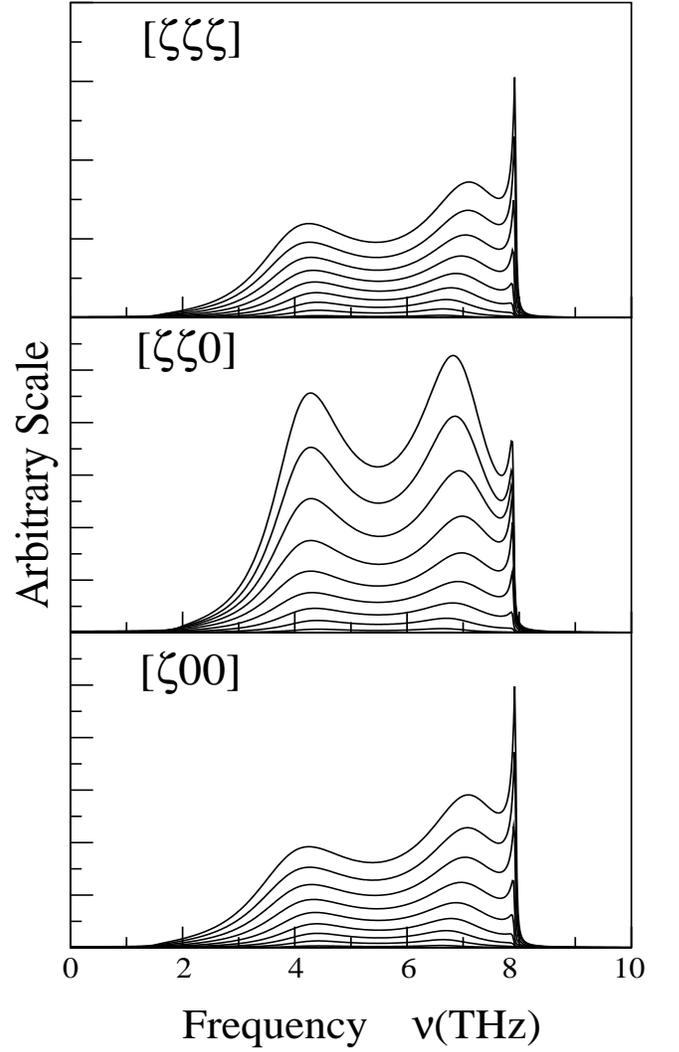}
\caption{The incoherent scattering cross sections in different directions for $Ni_{55}Pd_{45}$ alloy with $\Phi_{Ni-Pd}=0.9\ \Phi_{Ni-Ni}\ $. In each of the different directions, the various curves indicate the cross sections for various $\zeta$ values starting from the lowest value to the edge of the Brillouin zone.}
\label{fig10}
\end{figure}

In Fig.(\ref{fig9}) we display the incoherent scattering cross sections [\ calculated from Eqn.(\ref{incoh})\ ] along the highest symmetry directions. In a particular direction, the different curves indicate the cross sections for various $\zeta$-points starting from $\zeta =0$ to $\zeta =1$ (in units of $2\pi/a$). A look at fig.(\ref{fig9}) immediately shows that the incoherent cross sections are very weakly dependent on ${\mbf q}$. It is the "$q^{\alpha}q^{\beta}$" factor in Eqn.(\ref{incoh}) which weights up the cross sections as we go on increasing ${\mbf q}$-points. These results are in accordance with the arguments of Nowak and Dederichs \cite{nd} and Yussouf and Mookerjee \cite{my}. They have also mentioned in their paper that, the weak ${\mbf q}$-dependence of incoherent scattering cross section arises because of it's strong similarities with the self energy diagram which is itself a short range matrix due to it's dependence on irreducible diagrams and hence vary rather weakly with ${\mbf q}$. Kamitakahara and Brockhouse also found a similar qualitative features for the coherent and incoherent cross sections in their Inelastic neutron scattering measurment.

Fig.(\ref{fig10}) shows the same Incoherent scattering cross sections for $Ni_{55}Pd_{45}$ alloy but with a different parametrization of force constants. Here we keep $\Phi_{Ni-Ni}^{\alpha\beta}$ and $\Phi_{Pd-Pd}^{\alpha\beta}$ the same as those of the pure materials (same as in Fig.\ref{fig9}) and reduced the $\Phi_{Ni-Pd}^{\alpha\beta}$ below the $\Phi_{Ni-Ni}^{\alpha\beta}$ by an $\alpha\beta$-independent factor as $\Phi_{Ni-Pd}^{\alpha\beta}~=~0.9\ \Phi_{Ni-Ni}^{\alpha\beta}$. However in Fig.(\ref{fig9}) it was $\Phi_{Ni-Pd}^{\alpha\beta}~=~0.7\ \Phi_{Ni-Ni}^{\alpha\beta}$. One can easily see the variation in shape of the cross section as the input parameter varies. It is because of this reason a prior information about the species dependence of the force constants is very important. Our approach here made no attempt to obtain the input parameters themselves from first principles, but rather resorted, as others did earlier, because the aim of this work is to establish the augmented space block recursion as a computationally fast and accurate method for the cross section calculation in context of phonon excitations in random alloys. Our future endeavor would be to rectify this, and attempt to obtain the dynamical matrix itself from more microscopic theories, so that we can have a unique shape of the cross sections unlike the present case.

\subsection{${\bf Ni_{50}Pt_{50}}$ alloy : Strong mass and force constant disorder}

\begin{figure}
\centering
\includegraphics[width=8.5cm,height=15cm]{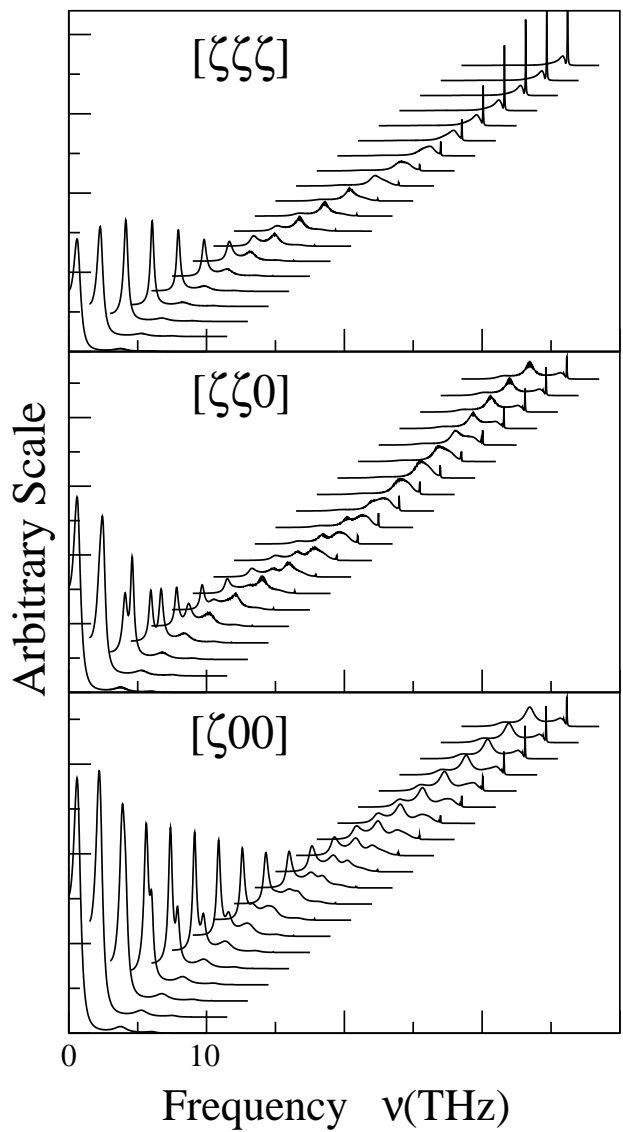}
\caption{The coherent scattering cross section in different directions for $Ni_{50}Pt_{50}$. In each of the different directions, the various curves indicate the cross sections for various $\zeta$ values starting from the lowest value to the edge of the Brillouin zone.}
\label{fig11}
\end{figure}
In this section we shall apply our formulation to NiPt alloys where both kinds of disorders are predominant. The mass ratio $m_{Pt}/m_{Ni}$ is 3.3 (quite large compared to that in NiPd system) and the force constants of Pt are on an average 55\% larger than those in Ni. For a list of general properties of fcc Ni and Pt, we refer the reader to article \cite{alam}. Tsunoda {\it et.al.} \cite{Tsunoda} investigated $Ni_{1-x}Pt_{x}$ by Inelastic neutron scattering and compared their observations with the CPA. Here for illustration, we have considered $x=0.5$ only because that makes it a concentrated alloy and the failure of the CPA was, qualitatively more prominent at this concentration. Because of the large mass and force constant differences, the effect of disorder in NiPt alloy is dramatic, such as the appearence of sharp discontinuities (or split bands) observed in the dispersion and line width \cite{alam}. The present theoritical investigation also found this kind of peculiar behaviour in the inelastic scattering cross sections. 

In Fig.(\ref{fig11}), we display the inelastic coherent scattering cross sections along the highest symmetry direction.As before, In a particular direction the different curves indicate the cross sections for various $\zeta$-values. For the sake of simplicity, we have used the same parametrization of masses and force constants as used in our previous paper \cite{alam}. 

In $Ni_{50}Pt_{50}$ alloy, the coherent scattering cross sections show few extra features. Even in the $[\zeta 0 0]$ direction, the cross section becomes well seperated double peaks along with weakly defined peak in between in the region from $\zeta = 0.68$ to the zone boundary. The occurence of such a weakly defined peak is due to the inclusion of force constant disorder explicitly in our formulation. It is also clear from Fig.(\ref{fig11}) that there exists no appreciable peak intensity below 3.7 THz for $\zeta \ge 0.68$ in all the three symmetry directions. Tsunoda also found the same structure below 3.5THz for $\zeta \ge 0.7$. However CPA predicted the lower frequency peak to exist for all the $\zeta$-values.

For smaller $\zeta$-values, the lower frequency peaks are sharper than the high frequency ones, but the intensity of former decreases significantly with increasing wave vector however the latter gets sharper, in all the three symmetry directions. The phonon peaks are well defined for smaller ($\zeta \le 0.38$) and higher $\zeta$-values, but no well defined peaks were observed for intermediate values of $\zeta$, presumably due to extreme line broadening. This kind of qualitative features has also been observed experimentally by Tsunoda {\it et. al.} for longitudnal branches. This feature is more transparent in the $[\zeta \zeta 0]$ direction [\ fig.(\ref{fig11})] where the low frequency peak gets broadened and becomes more asymmetric in the region between $\zeta = 0.4$ to $\zeta = 0.78$.

\begin{figure}
\centering
\includegraphics[width=8.5cm,height=14cm]{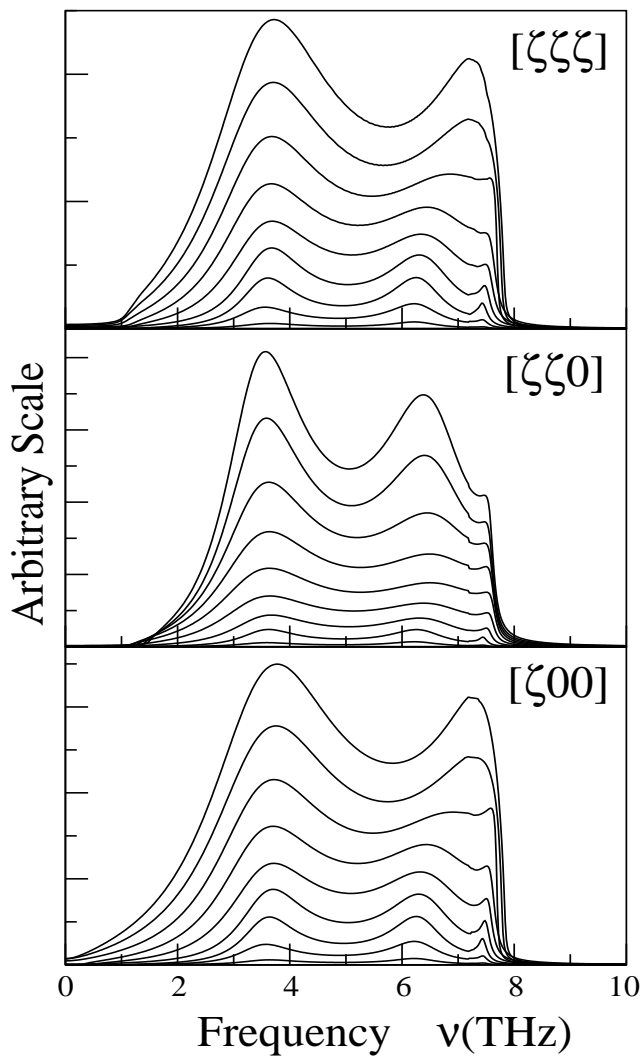}
\caption{The incoherent scattering cross section in different directions for $Ni_{50}Pt_{50}$. In each of the different directions, the various curves indicate the cross sections for various $\zeta$ values starting from the lowest value to the edge of the Brillouin zone.}
\label{fig12}
\end{figure}

The incoherent scattering cross sections (given by Eqn.\ref{incoh}) for $Ni_{50}Pt_{50}$ alloy along the highest symmetry directions are shown in Fig.\ref{fig12}. The weak $\mbf q$-dependence of the cross section is obvious from the figure. The intensities for various ${\mbf q}$'s in a particular direction have nearly an approximate $q^{2}$ dependence because of the factor '$q^{\alpha}q^{\beta}$' in Eqn.(\ref{incoh}).

\begin{figure*}
\centering
\includegraphics[width=17.0cm,height=6cm]{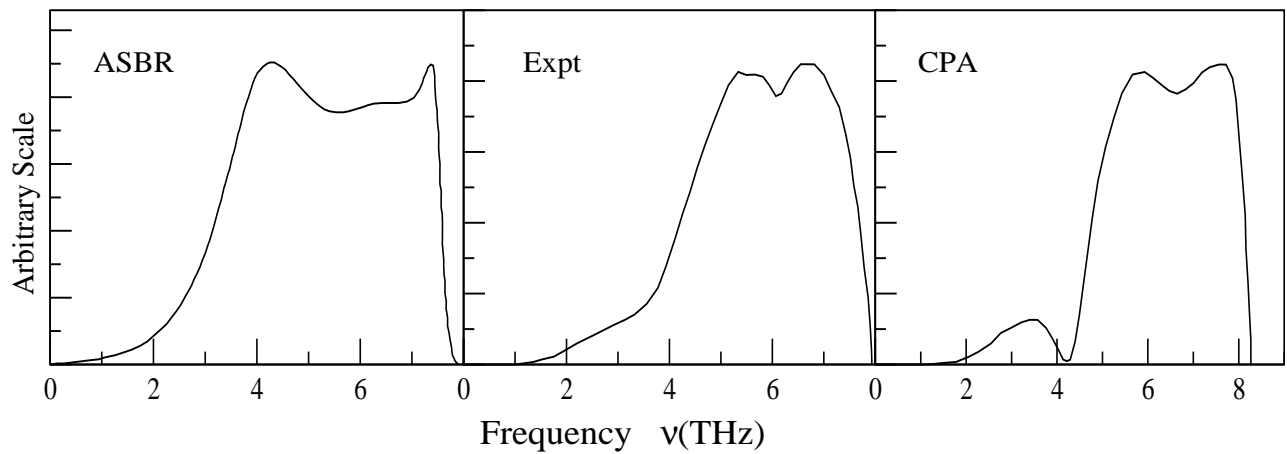}
\caption{The incoherent scattering cross section for $Ni_{50}Pt_{50}$. The left panel, middle panel and right panel display the block recursion result, experimental curve and the CPA result respectively. }
\label{fig13}
\end{figure*}
In Fig.(\ref{fig13}), we compare our results for the incoherent scattering cross section with those of the CPA and the experiment \cite{Tsunoda}. Here the left, middle and right panel displays the augmented space block recursion (ASBR) result, experimental curve and the CPA result respectively. In the CPA result we can observe a dip at the frequency corresponding to the phonon band gap observed in the dispersion curves. This suggests a split band behaviour which clearly seperates the Pt-contribution in the low frequency region from the Ni-contribution in the high frequency region, because the low frequency region is dominated by the Pt-atom (heavier atom) having much lower incoherent scattering length than Ni [\ the incoherent scattering length for Pt is 0.1 while that of Ni is 4.5\ ]. This kind of spurious gap however is not observed in the recursion result, because the CPA results are based on the mass fluctuations alone, ignoring the off-diagonal and environmental disorder arising out of the dynamical matrix. On the other hand, by incorporating the force constant disorder as is done in the bolck recursion, we get rid of this spurious gap and obtain rather a good agreement with the experimental results. The overall qualitative behaviour is similar. In addition the phonon band edges in the recursion results are very close to the experimental ones. The recursion finds the right band-edge at $\simeq 7.91THz $, Tsunoda finds this band-edge experimentally at $\simeq 7.93 THz $ while the CPA gives a rather higher value of $\simeq 8.267$ THz.

\section{Conclusions}

We have presented a straightforward and tractable formulation for the seperation of total intensity of thermal neutron scattering from disordered alloys into a coherent and an incoherent part. The use of the augmented space to keep track of the configuration of the system has made the formalism simple yet powerful. In essence, the splitting is identical to that introduced by Nowak and Dederichs \cite{nd} within a Yonezawa-Matsubara diagram technique except that it has been done {\sl exactly} without taking any recourse to mean-field like approximation. Unlike the method proposed by Yussouff an Mookerjee \cite{my}, where the diagram technique was exceedingly difficult to generalize into even a small cluster CPA, the augmented space block recursion proved to be simpler to apply the formalism on realistic random alloys. The technique takes into account fluctuations in masses, force constants and scattering lengths of the individual nuclei. The environmental disorder arising out of the force constant sum rule has also been incorporated before averaging. The approximation involving termination of matrix continued fraction expansion of the Green matrix retains the essential Herglotz analytic properties of the diagonal Green's function. We have applied the method to NiPd and NiPt alloys. In $Ni_{55}Pd_{45}$, we have demonstrated that mass disorder plays the prominent role. In addition our coherent scattering cross section enable us to understand the effect of small contribution of the off-diagonal elements of Green matrix. The results on $Ni_{50}Pt_{50}$  alloy however demonstrate the prominence of force disorder even in a case where the mass ratio is $\simeq$ 3. Our results agree well both with the coherent and the incoherent scattering experiments, where as the CPA fails both qualitatively and quantitatively. We propose the technique as a computationally fast and efficient method for the study of inelastic neutron scattering in disordered systems. Our approach here had no prior information about the species dependence of the force constants, but rather choose a set of force constants intutively as others did earlier. A better understanding of the role of disorder in the lattice dynamics of random alloys could be achieved with prior information about the force constants. These could be obtained from more microscopic theories (\ e.g. the first principles calculation on a set of ordered alloys.)

\end{document}